\documentstyle[12pt]{article}

\def\l#1{\label{#1}}
\def\be{\begin{equation}}
\def\ee{\end{equation}}
\def\bea{\begin{eqnarray}}
\def\eea{\end{eqnarray}}
\def\ot{\otimes}
\def\={\; = \;}
\def\G{\Gamma}
\def\T{\Theta}
\def\al{\alpha}
\def\ba{\beta}
\def\ka{\kappa}

\def\ti#1{{\tilde{#1}}}
\def\r{\rho}

\def\ga{\gamma}
\def\bt{\bullet}
\def\z{\hbox{\bf Z}}
\def\pt{\partial}
\def\a{{\cal A}}
\def\R{{\cal R}}

\def\sz{\hbox{$\hbox{\footnotesize Z}_{\hbox{\tiny 2}}$}}

\begin{document}
\title{Gravity From Noncommutative Geometry}
\author{Andrzej Sitarz \thanks{Partially supported by
KBN grant 2 P302 168 4} \thanks{E-mail: sitarz@if.uj.edu.pl} \\
Department of Field Theory \\
Institute of Physics \\
Jagiellonian University \\
Reymonta 4, 30-059 Krak\'ow, Poland}
\begin{titlepage}
\vspace{2cm}
\vfill
\maketitle
%\vfill
\begin{abstract}
We introduce the linear connection in the noncommutative geometry
model of the product of continuous manifold and the discrete space of
two points. We discuss its metric properties, define the metric
connection and calculate the curvature. We define also the Ricci
tensor and the scalar curvature. We find that the latter differs from the
standard scalar curvature of the manifold by a term, which might be
interpreted as the cosmological constant and apart from that we find
no other dynamical fields in the model. Finally we discuss an example
solution of flat linear connection, with the nontrivial scaling dependence
of the metric tensor on the discrete variable. We interpret the obtained
solution
as confirmed by the Standard Model, with the scaling factor corresponding to
the
Weinberg angle.
\end{abstract}
\vfill
\sc \noindent TPJU - 1/94 \\
\sc January 1994
\vfill
\end{titlepage}

%\baselineskip=1.75\baselineskip
\section{Introduction}

The interpretation of the electroweak part of the
Standard Model as arising from the product of
continuous and discrete geometry has been a
major success of noncommutative geometry
\cite{NCG1}-\cite{JA2}. It allowed to explain the geometrical origins
 of the Higgs field and the symmetry breaking mechanism. It is
important that in that approach the discrete geometry
is on equal footing with the continuous one, thus
raising questions on the possible effect of the discrete
geometry on gravity \cite{GRA1}-\cite{KAS}.
In this paper  we would like to discuss briefly few important points of the
metric aspects of such a model. We introduce the
analogue of metric tensor, the group of transformations of
the module of one-forms and we present the symmetry group
of the metric. Next, we introduce the linear connection and
discuss the construction of the curvature, Ricci tensor and
the scalar curvature for such model, which would be the starting
point for the investigation of Einstein-Hilbert gravity action.
Finally, we give an example of the solution with vanishing curvature,
which is nontrivial, i.e. the metric tensor  is scaled if we change from one
point of the discrete space to another.

\section{Differential Geometry on $M \times \z_2$}

In this section we shall briefly outline the construction of
the noncommutative differential geometry on a product of the
continuous manifold $M$ and the discrete two-point
space $\z_2$ (for details see \cite{JA1}). Our starting point is the algebra
$\a$ of
complex-valued functions on $M \times \z_2$, with the pointwise
addition and multiplication. This is a simpler model than the one
arising from the Standard Models, as in the latter case the  algebra consists
of functions complex-valued on one and quaternionic valued on the other
point.

Let $\Omega(M)$ denote the differential algebra on $M$ and
$d_M$ be the external derivative on $\Omega(M)$.
The differential algebra on $\z_2$ is constructed as follows.
Let $\Omega^1(\z_2)$ be a free bimodule over the algebra $A$ of  functions on
$\z_2$
generated by a given one-form $\chi$. The left- and right- multiplication by
the elements
 of $A$ are related by a simple rule:
\be
f \chi \= \chi \R(f), \label{r1}
\ee
where $R$ is the morphism of the algebra $\a$, which corresponds to the
interchanging of the two points:
\bea
\left( \R(f) \right)(x,+) & = & f(x,-), \\
\left( \R(f) \right)(x,-) & = & f(x,+),
\eea
the external derivative $d_{\sz_2}$ acts on $\a$ in the following
way:
\be
d_{\sz_2} f \= \chi \pt f \; \equiv \; \chi \left( f - \R(f) \right).
\label{r2}
\ee
The higher order forms are the tensor products (over $A$) of one-forms,
and the external derivative $d_{\z_2}$ acts on $\chi$ as follows:
\be
d_{\sz_2} \chi \; = \; 2 \chi \otimes \chi.
\label{r3}
\ee
The conjugation rule, which (graded) commutes with the external
derivative is as follows:
\be
\chi^\star \; = \; - \chi.
\label{r4}
\ee
Now, the differential algebra on $M \times \z_2$ is just the tensor product
of  $\Omega(M)$ and $\Omega(\z_2)$. If we assume for simplicity that
$\Omega^1(M)$ is a free module (which is equivalent to the triviality
of the tangent bundle TM) we shall have the module $\Omega^1(M\times \z_2)$
to be a free five-dimensional module over $\a$. The rules for the
multiplication
of forms in $\Omega$, which we shall denote by $\bt$, result from the rules of
multiplication in $\Omega(M)$ and the properties of the product:
\bea
dx^\mu \bt dx^\nu \= - dx^\nu \bt dx^\mu, \\
\chi \bt dx^\mu \= - dx^\mu \bt \chi.
\eea
Let us point out here that up to this point the construction closely
resembles the usual Kaluza-Klein theory, the only difference being
the specific bimodule structure of the differential algebra over
the additional (discrete) dimension. However, in the next section
we shall see that the consequences of this feature are profound and
make this theory much different from the standard Kaluza-Klein
approach.

\section{The Metric}

In the standard differential geometry the metric tensor can be
defined as a bilinear functional from the product of two copies
of $\Omega^1$ to the algebra of functions. We shall generalise this
definition to the noncommutative case and take the metric $g$ to be
a middle-linear functional:
\be
g: \Omega^1(M) \times \Omega^1(M)  \; \to \; \a,
\ee
where the middle-linearity is defined as follows:
\be
g(a \omega_1 b, \omega_2 c) \; = \; a g(\omega_1, b\omega_2) c,
\ee
for all  $\omega_1,\omega_2 \in \Omega^1$ and $ a,b,c \in \a$.
Additionally we shall require the hermicity of the metric, i.e.:
\be
g(\omega_1, \omega_2) \; = \; \left( g(
\omega_2^\star, \omega_1^\star ) \right)^\star.
\ee
Let us notice here that the expressions $g(\omega,\omega^\star)$
and $g(\omega^\star, \omega)$ are different and cannot be equal
unless the metric is degenerate and $g(\chi, \chi) = 0$.

Our last requirement is that there exists a trace (integration)
on the algebra $\a$ such the following condition are satisfied:
\be
\int g(\omega^\star, \omega) = \int g(\omega, \omega^\star),
\ee

If we impose all these conditions in the considered example
we obtain the following restrictions on the parameters of the metric:
\begin{eqnarray}
g(dx^\mu, dx^\nu) \; = \;  g^{\mu\nu} \l{me1a} \\
g^{\mu\nu} \; = \; \bar{g}^{\nu\mu} \; = \; g^{\nu\mu}, \l{me1b} \\
g(dx^\mu, \chi) \; = \; 0 \; = \; g(\chi, dx^\mu), \l{me1c} \\
g \; = \; g(\chi, \chi) \; = \; \bar{g}, \label{me1d} \\
g  \=  \R(g)  \l{sym1}.
\end{eqnarray}
The metric on the product space $M \times \z_2$ is therefore completely
determined by two 'standard' real metric tensors (each for the separate copy of
$M$)
and a scalar real function.

\section{Symmetries}

Before we begin the discussion of  linear connections in the considered
model let us say few words about its symmetries. As in the
 standard differential geometry the group of diffeomorphism
of the manifold corresponds to the conjugation preserving
automorphisms of its differential algebra, which commute with
the external derivative, our attention here shall be focused
on such a group. It is easy to demonstrate that in the particular case the
group of automerphisms of $\a$  is simply:$(Diff(M) \times Diff(M) ) \times
\z_2$
as we can have two separate diffeomorphisms on each copy of $M$
as well as we can interchange the points of the discrete space $\z_2$.

Now let us have look at the possible linear morphisms of
the $\Omega^1$, which preserve the conjugation. Let us
choose the left-linear morphisms. From the
rules (\ref{r1},\ref{r4}) we obtain that $A(v^\star) = A(v)^\star$
only if such morphism cannot mix the continuous and discrete
degrees of freedom:
\begin{eqnarray}
A(dx^\mu)  & \; = \; & A^\mu_\nu dx^\nu, \label{s1} \\
A(\chi)    & \; = \; & a \chi \label{s2}
\end{eqnarray}
where $A^\mu_\nu$ must be real and $a$ is such an element of
the algebra $\a$ that
\be
\R(a) \= \bar{a}, \label{s3}
\ee
Now, let us check, which of these morphisms (\ref{s1},\ref{s2})
are metric preserving, i.e. $g(u,v) = g(A(u),A(v))$ for
every one-forms $u,v$. It appears that the conditions are:
\begin{eqnarray}
A^\mu_\rho g^{\rho\sigma} A^\nu_\sigma & = & g^{\mu\nu}, \label{sm1} \\
\alpha \bar{\alpha} g & = & g, \label{sm2}
\end{eqnarray}
and we see that $A^\mu_\nu$ must be a function
valued in the Lorentz group whereas $\alpha$
must be unitary.

\section{Linear Connection}

In the construction of the linear connection in the examined
model we shall follow the procedure of translating the definitions
from the 'standard' differential geometry to the
noncommutative situation. We shall begin here with the introduction
of the covariant derivative $D$. We take $D$ as a mapping from
$\Omega \ot \Omega^1$, which satisfies:
\be
D(\omega \otimes u) \; = \; d\omega \otimes u
+ (-1)^{\hbox{\footnotesize deg} \omega} \omega \bullet D(u).
\l{lc1}
\ee
for any $\omega \in \omega$ and $u \in \Omega^1$.
Let us observe that any two such operators $D$ and $D'$ differ by
a (graded) left-linear operator over $\Omega \otimes \Omega^1$:
\be
(D-D') (\omega \otimes u) \= (-1)^{\hbox{\footnotesize deg} \omega}
\omega \bullet (D-D')u.
\l{lc2}
\ee
Due to the definition (\ref{lc1}) $D$ is completely determined on the
basis of one-forms, thus it is sufficient to calculate:
\bea
D(dx^\mu) \= \G^\mu_{\rho\sigma} dx^\rho \ot dx^\sigma
+ \G^\mu_\rho \chi \ot dx^\rho,
\l{lc5a} \\
D(\chi) \= \G \chi \ot \chi + \G_\mu dx^\mu \ot \chi.
\l{lc5b}
\eea
We have assumed here that the left-linear operators acting on $\Omega^1$
are generated by the star preserving morphism of $\Omega^1$ and therefore
due to (\ref{s1}-\ref{s3}) the coefficients must satisfy the following
relations:
\bea
\G^\mu_{\nu\r} & = & \bar{\G}^\mu_{\nu\r}, \l{re1} \\
\G^\mu_\nu & = & \bar{\G}^\mu_\nu, \l{re2} \\
\R\left( \G_\mu \right) & = & \bar{\G}_\mu, \l{re3} \\
\R\left( \G \right) & = & \bar{\G}. \l{re4}
\eea
Now, let us require that the above introduced linear connection
is the metric one, which now we might formulate in the following way:
\be
d \left( g(u,u^\star) \right)  \= \tilde{g}(Du,u^\star) +
\tilde{g}(u, (Du)^\star), \l{meco}
\ee
where $u$ is an arbitrary one-form and $\tilde{g}$ is the extension of the
metric $g$ to the forms having values in a module, the latter being
again $\Omega^1$ in our case:
\be
\tilde{g}(u_1 \ot u_2, v) \= u_1 g(u_2,v). \l{meco2}
\ee
The equation (\ref{meco}) together with the conditions (\ref{re1}-\ref{re4})
lead to the following set of constraints on the linear connection:
\bea
{\pt}_\r g^{\mu\nu} & = & \G^\mu_{\r\kappa} g^{\kappa\nu} + \G^\nu_{\r\kappa}
g^{\mu\kappa}, \l{lcc1} \\
\pt g^{\mu\nu} & = &  \R(\G^\mu_\kappa) g^{\kappa\nu} - \R(g^{\mu\kappa})
\G^\nu_\kappa, \l{lcc2} \\
\pt_\mu g & = & g( \G_\mu + \bar{\G}_\mu), \l{lcc3} \\
{\pt} g & = & 0, \l{lcc4}
\eea
Let us observe that (\ref{lcc4}) simply repeats the condition (\ref{sym1}),
which we have obtained earlier. The first condition (\ref{lcc1}) gives us
the usual relation between the Christoffel symbols and the derivatives of
the metric tensor.

As a next step we shall present the transformation laws of all
introduced coefficients of the curvature induced by the star
preserving automorphisms of $\Omega^1$, which we have discussed
earlier. If $A$ and $a$ denote the transformations of the
continuous and discrete degrees of freedom respectively, and
$\ti{A}$ is $A^{-1}$, we shall have:
\bea
\G^\al_{\ba\ga}  & \to &
\G^\mu_{\ba\r} A^\al_\mu \ti{A}^\r_\ga
+ \pt_\ba (A^\mu_\ka) \ti{A}^\ka_\ga, \l{trl1} \\
\G^\al_\ba & \to & \G^\mu_\nu
 A^\al_\mu \R(\ti{A}^\nu_\ba)
- \left( A^\al_\ka - \R(A^\al_\ka) \right) \R( \ti{A}^\ka_\ba), \l{trl2} \\
\G_\mu & \to & \frac{1}{\bar{a}} \left( \G_\ka + \pt_\ka a \right), \l{trl3} \\
\G & \to & \frac{1}{\bar{a}} \left( \bar{a} - a + a\G \right). \l{trl4}
\eea
One may easily verify that the constraints (\ref{lcc1}-\ref{lcc4}) are
invariant under these transformations. Let us observe that although
$\G^\mu_\nu$ is not a tensor with respect to the change of coordinates
on both copies of $M$, we may find that $\T^\mu_\nu = \delta^\mu_\nu -
\G^\mu_\nu$ transforms like a tensor:
\be
\T^\mu_\nu \; \to \; \T^\al_\ba  A_\al^\mu \R(\ti{A}_\nu^\ba),
\ee
Similarly, we may also introduce $\T = 1 - \G$, which transformation
law is again simpler as the one for $\G$:
\be
\T \; \to \; \frac{1}{\bar{a}} \T a.
\ee
Using the newly introduced variable $\T^\mu_\nu$ we may rewrite the
constraint (\ref{lcc2}) as
\be
\R(\T^\mu_\ka) g^{\ka\nu} \= \R(g^{\mu\ka}) \T^\nu_\ka. \l{ncon}
\ee

In the standard differential geometry there exists exactly one linear
metric connection, which has a vanishing torsion. So far in our
considerations we have neglected the torsion and now we have to
explain why this point cannot be transferred as it stands to the
noncommutative case. Originally, the reason that
one may restrict oneself to the connections with no torsion
comes from the geometric picture. It appears that for every
linear connection there exists one with vanishing torsion, which
has the same geodesics. As in the noncommutative case we do not
have such a picture, we have no motivation for imposing the same
condition in general. Naively, since we may interpret torsion
as the following map:
\be
T: \; \Omega^1 \ni \omega \; \to  (\pi \circ D - d) \omega \in \Omega^2,
\ee
where $\pi$ is the projection $\pi: \Omega^1 \ot \Omega^1 \to \Omega^2$,
by imposing that the torsion vanishes we shall obtain that $\G^\mu_\nu$
and $\G_\mu$ must vanish and $\G=2$. Additionally,
from the constraint (\ref{lcc2}) we shall have that
$\pt g^{\mu\nu} = 0$, which would mean that the metric on both copies
of $M$ shall be the same and the resulting theory of
linear connections would be the same as for the single manifold $M$.

In our case, however, we shall only require that the torsion restricted
to the continuous degrees of freedom vanishes, which is equivalent to
the symmetry of the Christoffel symbols:
\be
\G^\mu_{\r\nu} \= \G^\mu_{\nu\r}.
\ee

Finally, let us come to the construction of curvature. Similarly as
in the ordinary differential geometry we define the curvature as the
square of the covariant derivative $R =D^2$, which then is a left-linear
operator on $\Omega \ot \Omega^1$:
\be
R(\omega \ot u) \= \omega R(u), \l{cur1}
\ee
so in order to determine $R$ it is sufficient to calculate its action
on the basis of one-forms:
\bea
R(dx^\mu) & = &
dx^\al \bt dx^\ba \left( \pt_\al \G_{\ba\r}^\mu - \pt_\ba \G_{\al\r}^\mu
- \G_{\al\ka}^\mu \G^\ka_{\ba\r} +  \G_{\ba\ka}^\mu \G^\ka_{\al\r} \right)
\ot dx^\r \l{cu1} \\
& + & dx^\al \bt \chi \left( \R( \pt_\al \T^\mu_{\r})
+ \R(\G^\mu_{\al\ka} \T^\ka_\r) - \R(\T^\mu_\ka) \G^\ka_{\al\r} \right)
\ot dx^\r \l{cu2} \\
& + & \chi \bt \chi \left( \delta^\mu_\al - \T^\mu_\ka
\R(\T^\ka_\al) \right) \ot dx^\al, \l{cu3}
\eea
and
\bea
R(\chi) & = & dx^\al \bt dx^\ba \left( \pt_\al \G_\ba
- \pt_\ba \G_\al \right) \ot \chi \l{cu4} \\
& + & dx^\al \bt \chi \left(
\R(\pt_\al \T) + \R(\T \G_\al) - \R(\T) \G_\al \right) \ot \chi \l{cu5} \\
& + & \chi \bt \chi \left( 1 - \T \R(\T) \right) \ot \chi. \l{cu6}
\eea
The first component of the curvature (\ref{cu1}) is just the standard
Christoffel-Riemann curvature tensor (for both copies of $M$),
whereas the rest is the result of the existence of additional structure
of discrete geometry. Let us notice that (\ref{cu4}) vanishes if we
take into account the equation (\ref{lcc4}).

We shall briefly discuss an example of a nontrivial solution with the
vanishing curvature in the next section and now we shall proceed with
the definition of the generalisation of the Ricci tensor and the scalar
curvature. We define the Ricci tensor $R_c$ as the trace of $R$, treated
as a map $R: \Omega^1 \to \Omega^1 \bt (\Omega^1 \ot \Omega^1)$. In this
approach $R_c$ is simply an element of $\Omega^1 \ot \Omega^1$ and we
have:
\bea
R_c & = &
 dx^\ba \left( \pt_\mu \G_{\ba\r}^\mu - \pt_\ba \G_{\mu\r}^\mu
- \G_{\mu\ka}^\mu \G^\ka_{\ba\r} +  \G_{\ba\ka}^\mu \G^\ka_{\mu\r} \right)
\ot dx^\r \l{rc1} \\
& + & \chi \left( \R( \pt_\mu \T^\mu_{\r})
+ \R(\G^\mu_{\mu\ka} \T^\ka_\r) - \R(\T^\mu_\ka) \G^\ka_{\mu\r} \right)
\ot dx^\r \l{rc2} \\
& - &  dx^\al \left( \R(\pt_\al \T) + \R(\T \G_\al)
- \R(\T) \G_\al \right) \ot \chi \l{rc3} \\
& + &  \chi \left( 1 - \T \R(\T) \right) \ot \chi \l{rc4}
\eea

The scalar curvature $R_s$ is the contraction of $R_c$ with the
metric $g$ (the latter treated as a map from
$\Omega^1 \ot \Omega^1 \to \a$. Due to the special properties of
$g$ we obtain the scalar curvature the same as in the standard
differential geometry with an additional cosmological constant term:
\be
R_s \= g^{\ba\r} \left( \pt_\mu \G_{\ba\r}^\mu - \pt_\ba \G_{\mu\r}^\mu
- \G_{\mu\ka}^\mu \G^\ka_{\ba\r} +  \G_{\ba\ka}^\mu \G^\ka_{\mu\r}
\right) + g (1 - \T \R(\T)).
\ee
Let us point out that the remarkable feature that the scalar curvature
has no other dynamical variables that the restriction of the metric
to the continuous coordinates $g^{\mu\nu}$. The other field $\T$, which
appears in the above expression, has no dynamics and therefore plays
only a role of the scale factor of the cosmological constant.

\section{Example solution}

In this section we shall briefly discuss an example of a non-trivial
solution of a flat linear connection, which we can compare with
the information obtained from the Standard Model.

Let us assume for simplicity that $\G_\mu=0$, $g=1$ and $\T = 1$. Then,
the whole information of the linear connection is in dynamical
variable $g^{\mu\nu}$ and the auxiliary one $\T^\mu_nu$.

Let us now take $\T^\mu_nu = \Psi \delta^\mu_\nu$, where $\Psi$
is an element of $\a$, which is constant on each copy of $M$. From
the vanishing of the component (\ref{cu3}) of the curvature we have
\be
1 \= \Psi(+) \Psi(-),
\ee
which allows us to write $\Psi(+)$ as $\tan \varphi$ and
$\Psi(-)$ as $\cot \varphi$ for some nonzero real number
$\varphi$. Next, let us us the constraint (\ref{ncon}), we obtain
then the following relation:
\be
( \tan  \varphi ) \R(g^{\mu\nu}) \= g^{\mu\nu} (\cot \varphi ),
\ee
which could be easily solved. If we define $h^{\mu\nu}$ as follows
\be
h^{\mu\nu} \= g^{\mu\nu} + \R (g^{\mu\nu}),
\ee
then we shall simply have:
\bea
g^{\mu\nu}(x,+)  & = & (\sin^2 \varphi) h^{\mu\nu}, \\
g^{\mu\nu}(x,-)  & = & (\cos^2 \varphi) h^{\mu\nu}.
\eea
Let us observe that it guarantees that the linear connection
$\G^\mu_{\al\r}$ are identical on both copies of $M$, thus,
if $h^{\mu\nu}$ is the flat Minkowski metric, we shall have
all components the curvature tensor (\ref{cu1}-\ref{cu6})
vanishing.

The obtained solution  explicitly demonstrates that
in the noncommutative extension of gravity on the product
of continuous and discrete geometry there are possible some
nontrivial solutions with vanishing curvature. What is more
intriguing, we can link this solution with the interpretation
of the Standard Model and the Weinberg angle \cite{JA1,JA2,JA3}.
It appears that the nature of the Weinberg angle can be reinterpreted in the
noncommutative geometry approach as the difference in scaling of
the flat metric tensor on the different copies of the Minkowski
space corresponding to left and right fermions. Now, if we look
at the above solution, we see that it is precisely what we have
derived as an example solution of flat metric on $M \times \z_2$.
Therefore we find an amazing agreement between the theoretical
solutions and the experimental facts, as explained by the
form of the Standard Model.

\section{Conclusions}

If the Standard Model gives us the correct (at least at the current energy
level)
description of the elementary particles we have strong reasons to believe that
at small distances the geometry of the world is more subtle that once though
and it involves the  discrete geometry. As we know what are its consequences
for the gauge theories it is reasonable to ask whether the resulting theory of
gravity would differ from the standard general relativity. So far, there have
couple of attempts to solve this problem \cite{GRA1}-\cite{KAS},
with various results. Here, we have demonstrated that although the linear
connection
has some auxiliary fields, none of them enters the scalar curvature and the
action in any other form
than the scaling factor. Therefore the resulting theory of gravity would be,
in general, similar to standard one, the only differences being the possibility
of separate solutions for $g^{\mu\nu}$ on each copy of $M$ and the
additional cosmological constant term in the generalised scalar  curvature.
Our results seem to agree with these derived in a recent paper by Kalau and
Waltze \cite{KAL}, where the gravity action was obtained by taking the Wodzicki
residue of the operator $D^{-2}$.

In this article we have explicitly derived an example of a solution with the
vanishing curvature, which has different metric on the separate copies of the
manifold $M$. The interpretation of the variable $\varphi$ as the Weinberg
angle agrees comes directly from the interpretation of the Standard Model
in the noncommutative-geometry framework. Therefore, this variable should be
treated rather a parameter connected with geometry than another free parameter
of the Standard Model. The possible consequences of this interpretation might
be profound, for instance, it should be verified whether the Weinberg angle
could
be generalised to be a dynamical variable.

As in this paper we have only touched the subject of the generalisation of
gravity
in noncommutative geometry, there remains a lot of interesting problems for
further research. The above mentioned problem of the Weinberg angle seems to be
the most intriguing one. One should also attempt to find other possible
solutions
and verify to what extent they might differ from the standard solutions.
Another
problem is the coupling between gravity and matter fields, which, in this
approach
could be well formulated.

\vspace{1.5cm}
\noindent{\bf Acknowledgements}\\[0.5cm]
I would to thank R.Coquereaux for his kind hospitality and helpful discussions
during my stay in Marseille.
\def\v#1{{ \bf  #1} }

\end{document}